# Are microbial *biosurfactants* actually only *surfactants*?

Niki Baccile[a,*]

[a] Sorbonne Université, Centre National de la Recherche Scientifique, Laboratoire de Chimie de la Matière Condensée de Paris, LCMCP, F-75005 Paris, France

* Corresponding author:
Dr. Niki Baccile
E-mail address: niki.baccile@sorbonne-universite.fr
Phone: +33 1 44 27 56 77

**Abstract**

The term *biosurfactants* refers to a complex mixture of metabolites with surface-active properties produced by specific microorganisms. However, nowadays trends moves towards isolation, screening and purifying single biocompatible, biodegradable biosurfactants with high commercialization potential. Current legislation limiting petrochemicals combined with environmentally concerned consumers did not only stimulate research and development but it also promoted large-scale production of this class of molecules. However, recent data recorded on single congeners question the actual pertinence of using the word "*biosurfactant*" associated to these molecules. By evaluating the accepted characteristics of *surfactants* and comparing them to the actual self-assembly and bulk properties in water of molecules traditionally called "*biosurfactants*", this opinion paper aims at showing that the term "*biosurfactant*" can be somewhat reductive when applied to specific individual compounds produced by fermentation. The use of a more generic term, like *bioamphiphile* could probably be more pertinent and appropriate for consideration in the future.

*This article is published within the context of the Special Issue on Biosurfactants at Current Opinion in Colloid and Surface Science*

**Biosurfactants today**

Fully bio-based surfactants are generally intended to describe molecules either prepared by chemical synthesis from bio-based synthons (e.g., alkylpolyglucoside esters) or by biological approaches such as extraction from plants (e.g., saponins), biocatalysis (alkylpolyglucosides from enzymatic catalysis) or fermentation using selected microorganisms, such as yeasts or bacteria.[1] Decades of research in this field has continued on the tradition of using the term "*biosurfactant*" solely associated with the products obtained using the fermentation method.[2–5] The term "*biosurfactant*" is then associated to the surface active properties of complex mixtures of metabolites produced by some fungi and bacteria. According to the most plausible hypotheses, microorganisms use them to solubilize hydrophobic substrates as carbon source, to adhere and release to/from surfaces and enhance interfacial mobility, for their antiseptic properties or as energy reserve.[6] More recently, these compounds have been more and more topic of research in the field of industrial microbiology, which has the goal of controlling their production, extraction, purification and, overall, industrialization through a fermentative process.

From a historical perspective, biosurfactants are described in the literature since the mid of the twentieth century: sophorolipids, obtained from the yeast *Starmerella bombicola*, were first described in 1968 by Tulloch,[7,8] cellobioselipids, produced by the fungus *Ustilago maydis* and known as ustilagic acid, were reported first in 1951[9] while rhamnolipids, produced by the bacteria *Pseudomonas aeruginosa*, are known since 1949.[10–12] Other typical biosurfactants are mannosylerythritol lipids, produced by *Candida antarctica*.[13] All of these are grouped under the category of glycolipids (Figure 1), as they all contain a mono or disaccharide as hydrophilic headgroup and a lipid, generally a fatty acid backbone as the hydrophobic moiety. Another category of such molecules is composed of lipopeptides, like surfactin (Figure 1) and iturin A, produced by *Bacillus subtilis*.[14,15] Interestingly, although some phospholipids and polymeric emulsifiers (e.g. emulsan) have historically been classified as biosurfactants in most related literature in this area,[2–4,16–18] the majority of research share in this field is mostly focused on glycolipids and lipopeptides.

Fermentation-produced glycolipids and lipopeptides is certainly justified by their foaming properties,[19–24] which is often observed during the biosynthesis process, as a consequence of their ability to lower the surface tension of water solutions which is described in most related studies in literature over the years.[1,3,4,25] The existence of a critical micelle concentration (CMC) is also a typical *surfactant* feature and equally reported in a number of scientific papers.[1,3,4,25] Their emulsification properties and actual use in existing commercial formulations of green detergents further justifies the use of the term "*biosurfactant*".[3,4,16,17,26,27] Last but not least, antimicrobial effects, generally explained by their membrane lysis action similarly to many other positively and negatively-charged surfactants, were long described.[28]

However, recent research in this field carried out in the past decade shows evidence that the aqueous phase behaviour of fermentation-produced glycolipids and lipopeptides is more complex if taken as a whole[1] and it may provoke a more thoughtful reasoning on the actual pertinence of employing the term "*biosurfactant*", which defines a well-specific property, that is of being a surface active agent.

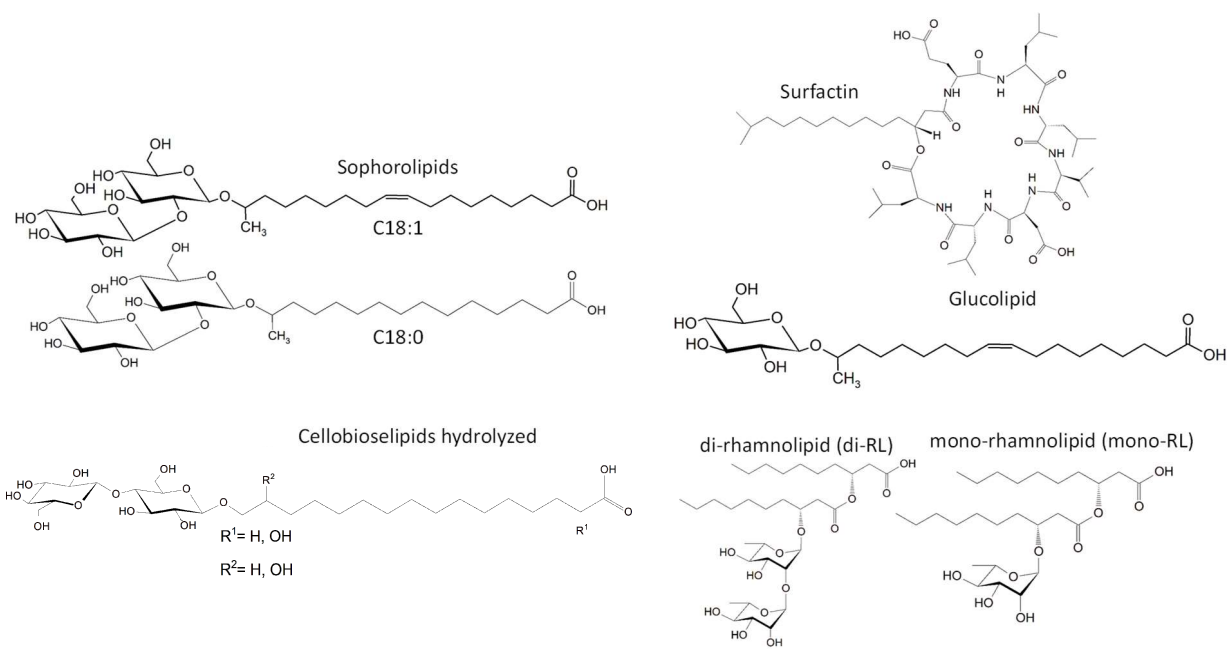

**Figure 1 – Classical set of microbially produced glycolipids and lipopeptides traditionally referred to in literature as "*biosurfactants*".**

**Surfactants *versus* amphiphiles**

As outlined in the Encyclopedia of Colloid and Interface Science (edition 2013, Page 1165),[29] surfactants are defined by four concomitant properties: 1- the ability to lower the surface tension of water, 2- the association into spherical aggregates, called micelles, beyond a 3- critical micellar concentration, CMC and finally, 4- micellization as a dynamic equilibrium process driven by the hydrophobic effect, generating a hydrophobic internal core and a hydrophilic shell containing the surfactant headgroups. A molecule characterized by a hydrophilic headgroup, and a lipophilic tail is typically given the general definition of an amphiphile, or amphipathic, and in many cases displays the characteristic behaviour of a surfactant in water above its Krafft point. The Krafft point, or Krafft temperature, is the temperature above which the surfactant shows a rapid increase in solubility with further increase of temperature (at the Krafft temperature, the solubility is equal to the CMC and an equilibrium occurs between hydrated surfactant solid, micelles, and monomers).[29] This is the case for many alkylammonium and alkyl sulfate salts, with a Krafft point broadly contained at around room temperature, between between 15°C and 30°C (e.g., 27-28°C for CTAB and 17°C for SDS).[29] A more pertinent definition of surfactants is the ability to foam and/or form oil-in-water emulsions.

However, not all amphiphiles share the features of surface active agents. Phospholipids are natural amphiphiles being the major constituent of the cell membrane. If they have a surface activity, they normally do not form micelles in the surrounding of room temperature, but stable bilayers, with reduced exchange dynamics. For these reasons, one of the most important works in the field of colloids science, *Theory of Self-Assembly of Hydrocarbon Amphiphiles into Micelles and Bilayers* by Israelachvili, Mitchell and Ninham[30] employs the term *amphiphile* in relationship to micelles- and bilayer-forming molecules. Nevertheless, one must acknowledge a certain ambiguity, as bilayer-forming molecules like phospholipids can also be considered as a subclass of zwitterionic surfactants.[29]

More peculiar are the case of amphiphiles which have a rigid steroid (bile salts, like lithocholic acid), steroid-like (e.g., glycyrrhizic acid), or more complex, or exotic (e.g., fluorenylmethyloxycarbonyl, FMOC), hydrophobic segments. The acidic form of this class of amphiphiles has a tendency to fibrillate, a trait often associated to gelation. The fibres are semi-crystalline and have an infinite aspect ratio.[31–33] Similarly, oleic acid, stearic acid or hydroxy stearic acid, just to cite some, do not belong to the domain of surfactants, but rather to the domain of oils and fats. Nevertheless, the solubility, micelle-forming and overall surfactant properties of all the molecules cited above may appear under alkaline conditions, when their corresponding salt is formed, as typically known for oleate and stearate metal salts.

**Recent finding on the self-assembly properties of microbial "biosurfactants"**

The study of the self-assembly behaviour of microbial biosurfactants is a relatively recent field of research. The first report was published in 1987 and showed the dual micellar (high pH, sodium salt) and vesicular (low pH, acidic form) nature of an aqueous solution of a mono- and di-rhamnolipids mixture (Figure 1).[34] This result has been confirmed by a number of additional studies.[35–38] Similar results were also reported for the lipopeptide surfactin (Figure 1), assembling into membranes at acidic pH[39,40] or mycosubtilin, assembling into fibers.[41] Sophorolipids, another well-known biosurfactant (Figure 1) typically obtained as a mixture of acetylated and non-acetylated acidic and lactonic forms bearing a C18:1 (oleil) backbone, can easily be converted into a non-acetylated acidic form.[42] The latter tends to form stable micellar solutions at a broad pH range.[43,44] As a matter of fact, this behaviour is probably related to the presence of minor amounts of other congeners, the complete removal of which is usually too tedious and expensive.[45] On the other hand, the behaviour of the non-acetylated acidic saturated (C18:0, Figure 1) form of sophorolipids is completely different than that all other biosurfactants. The acidic form of C18:0 sophorolipids fibrillates at low pH,[44,46,47] similarly to bile salts, while its sodium salt forms micellar solutions.[44,46] Similar results were observed for other microbially-produced glycolipids, like palmitic acid (C16:0) sophorolipids[48] and cellobioselipids.[49] The behaviour of bolaform sophorosides is also worth noting, as these molecules fibrillate and form hydrogels at room temperature, with a transition to micelles above about 30°C to 35°C.[50,51]

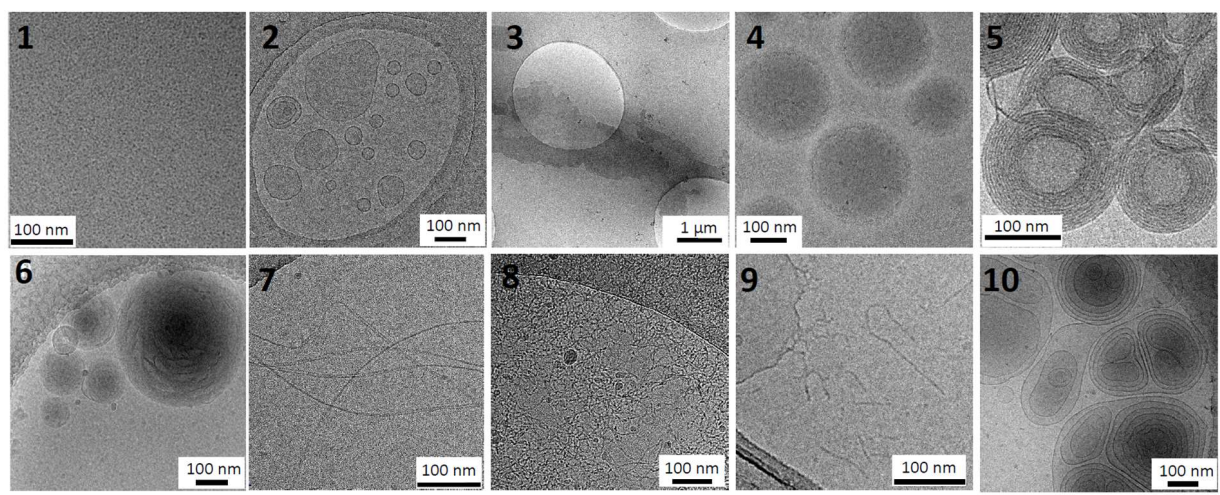

**Figure 2** - Cryo-TEM images recorded on diluted aqueous solutions of glucolipid C18:1 (Figure 1) prepared under different physicochemical conditions: 1) micelles, basic pH (counterion: sodium); 2) unilamellar vesicles, acidic pH; 3) lamellar precipitate, pH below 4; 4) complex coacervates, basic pH in the presence of a polycation; 5) multilamellar wall vesicles, acidic pH in the presence of a polycation after reducing the pH from basic to acidic; 6) aggregated vesicles, acidic pH in the presence of a polycation; 7) fibers, basic pH (counterion, $Ag^+$); 8) wormlike micelles, basic pH (counterion, $Fe^{2+}$); 9) wormlike micelles, pH between 6 and 7; 10) multilamellar wall vesicles, acidic pH after increasing pH from below 4. Reprinted from Ref. [52], with permission from Elsevier.

Even more peculiar is the behaviour of single-glucose lipids, both saturated, C18:0, and monounsaturated, C18:1 (Figure 1). Their acidic form at low pH assembles into single-layer membranes (may them be flat or curved)[44,49,53,54] while their sodium salt forms a water soluble micellar solution.[44,49] The deprotonated C18:1 glucolipid can even show a unique fibrillation and metallogelation behaviour in the presence of some monovalent and divalent cations ($Ca^{2+}$, $Ag^+$),[55] where cation-induced fibrillation is known for low-molecular weight gelators[56,57] but not for surfactants, which are known since long time to grow into large wormlike micelles when exposed to an excess of counterions in water.[58,59] In fact, the C18:1 glucolipid can display nine different phases in a limited span of pH at room temperature, depending on pH but also on the type of counterion and sense of varying pH.[52] The peculiar case of C18:1 glucolipid is sketched in Figure 2 and Figure 3, which show the typical cryogenic transmission electron micrographs and small angle X-ray scattering fingerprint of each phase, respectively. The overall self-assembly behaviour of most microbial glycolipids and lipopeptides has been reviewed by us recently,[1] while the specific behaviour of glucolipid C18:1[52] and of a number of chemo derivatives of sophorolipids has been reported elsewhere.[60]

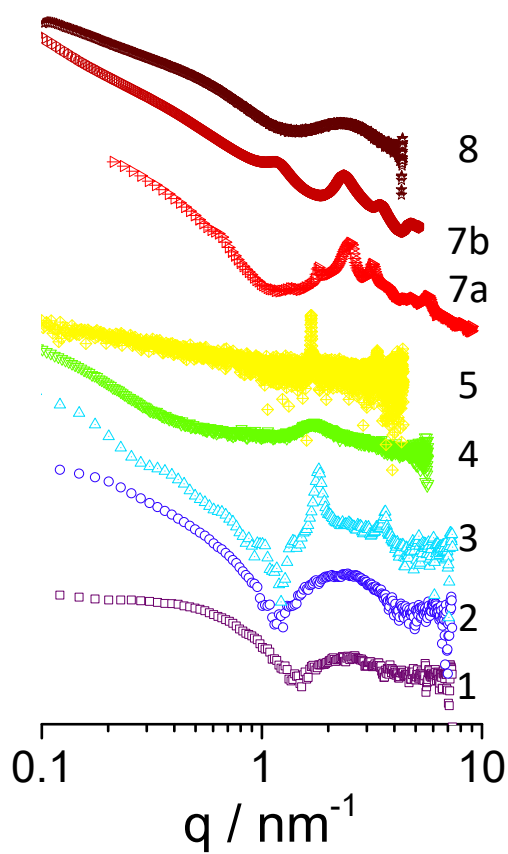

Figure 3 – Typical SAXS profiles recorded on glucolipid C18:1 (Figure 1) samples shown in Figure 2. Numbers indicate the same samples as in Figure 2, where 7a and 7b contain $Ca^{2+}$ and $Ag^+$, respectively, as counterions. Reprinted from Ref. [52], with permission from Elsevier.

**Bioamphiphiles**

The complex, parametric-dependent, self-assembly behaviour of most microbial glycolipids and lipopeptides is at the interface of surfactants, phospholipids but also bile salts, fatty acids and low-molecular weight gelators. In the best case scenario, the triple micelle-, membrane- and fibre-forming ability is observed across microbial glycolipids and lipopeptides, while in the most complex case, such variety is observed for a single molecule under comparable physicochemical conditions. This evidence casts doubt on the proper use of the word "*biosurfactant*". According to the definitions and descriptions given in the Encyclopedia of Colloid and Interface Science, when microbial glycolipids and lipopeptides assemble into vesicles or lamellae, the term "*biosurfactant*" is mildly appropriate, while in the case of fibrous assemblies, it would not be appropriate at all.

On the one hand, such multiphasic behaviour is generally observed for the neutral, generally acidic, form of this class of compounds, while the surfactant-like behaviour is generally observed for the corresponding deprotonated sodium salt.[1] However on the other hand, it is precisely the acidic form that is represented and addressed to as being a "*biosurfactant*" in the dedicated literature.[2–4,16–18] The question after all is quite simple: why does one employ the term oil, or fat or gelator, for families of molecules like fatty acids and bile salts in their uncharged, generally acidic, form, and the term *surfactant* for the uncharged acidic form of sophorolipids, rhamnolipids, glucolipids, cellobiose lipids or surfactin, when most of the times these fermentation-produced molecules behave as lipids and gelators, particularly at acidic pH values? Would the term *bioamphiphile*, be more generally appropriate and less restrictive?

The term *bioamphiphile* could refer to an amphiphile of biological origin, microbial in this case, and it has the advantage of qualifying a molecule for its structural features rather than its properties, for instance a surfactant or a gelator. The expression *microbial bioamphiphile* would then include microbial glycolipids and lipopeptides on the sole basis of their microbially-fermented origin and amphiphilic structures.

When addressing microbially-produced glycolipids and lipopeptides, the use of the term *amphiphile* is seldom used,[1,4,61–63] if compared to the term "*microbial biosurfactants*". One major reason is undoubtedly historical, as surface activity is classically associated to the mixture of metabolites produced by microorganisms; surface activity is also the property that is sought for when screening the production of new compounds by microorganisms.[6] Of course, surface activity is also the property that it has a major interest in view of replacing synthetic surfactants. Nevertheless, one of the first reviews in the field, published in 2002 by Prof. S. Lang, a microbiologist pioneer in the field of biosurfactants, was appropriately titled *Biological amphiphiles (microbial biosurfactants)*.[4] It seems that this review employs the term "*amphiphile*" specifically and explicitly towards microbially-produced glycolipids and lipopeptides for the first time, despite the fact that the field had already witnessed at least four other previously published review articles, which promoted the term "*microbial surfactants*".[2,3,16,25] However, the use of the term "*amphiphile*" was not specifically justified

*This article is published within the context of the Special Issue on Biosurfactants at Current Opinion in Colloid and Surface Science*

with respect to the specific properties of this class of molecules. The use of the term "*amphiphile*", and more precisely the expression *microbial bioamphiphile*, would be more accurate, following the example of Prof. Lang. The expression *microbial bioamphiphile* may also be less misleading and more appropriate, in view of the recent advances in the field.

It is generally accepted the fact that the surface activity of microbial metabolites is never associated to a single compound, but to a naturally-selected formulation of molecules.[64,65] Individual molecules in the formulation may actually have poor, if not any surface activity at all. This is the case of lactonic sophorolipids, the most abundant component (about 60-80%) in the classical raw mixture of lactonic/open acidic forms of this family of compounds.[65] Lactonic sophorolipids are more insoluble and with a tendency to form membrane structures, instead of micelles.[66] In this regard, its surfactant properties are arguable and use in surfactant-classics applications unlikely.[67] On the other hand, the open acidic form of sophorolipids can certainly be considered as a surfactant, but its interfacial performances are poor.[68] Interestingly, it was shown that synergy between the lactonic and open acidic forms have better foaming and interfacial properties,[67] a fact, which could help understanding the reason why mixtures of these congeners are produced in nature. A similar observation could be done for rhamnolipids, of which the natural formulation, about 50% of mono- and di-ramnose lipids, has optimized surface activity.[64] At neutral/acidic pH, mono-rhamnolipids tend to associate into vesicles, while di-rhamnolipids associate into into micelles, before precipitating into lamellar structures at pH below its pKa.[69,70] Finally, it is worth saying that many metabolites produced in much lower amounts, like the C18:0 or C16:0 forms of sophorolipids, do not have surface activity at all, as they are insoluble and fibrillate in their neutral form.[46,48] These facts suggest that biological evolution has engineered optimal mixtures of metabolites to enhance surface activity and to promote properties like emulsfication, adhesion or solubilisation, relevant for microbial survival.

Considering the fact that different congeners have different properties, a lot of research has been recently devoted to generate molecular variation, either through a genetic[71] or chemical[72,73] engineering approach, sometimes combined.[72] Engineered strains are able to produce both natural compounds, like acetylated or acidic sophorolipids,[24] and new-to-nature compounds, like glucolipids or sophorosides.[71,74] In the best case scenarios, yields are very good and selectivity high. Although not yet developed at industrial scales, a number of (chemically or genetically) engineered microbial metabolites certainly cannot be catalogued as surfactants. Nevertheless, they still have both scientific significance, especially in the space colloids and surface science,[52] and industrial interest.[75]

If one acknowledges that the field of microbial biosurfactants is evolving and rapidly expanding, it could be interesting, and most likely important, to improve the sharing of information and communication across communities, which for this field span from microbiology and genetic engineering to physical chemistry, colloids science, organic chemistry, process engineering and all the way to the private sector. The term "*biosurfactant*" implicitly implies that this class of compounds are surfactants of biological origin, with the obvious aim of replacing chemical surfactants in all applications where surfactants are employed, from detergency to cosmetics. However, as some single-molecule microbial glycolipids and lipopeptides, may them be obtained from wild type or genetically-modified microorganisms, may not display the expected surfactant-like properties under some conditions of use (e.g., low pH, room temperature, presence of specific salts), the expectations generated by the end-user (e.g., a private company) may be highly affected. Alternatively, an expression such as *microbial bioamphiphile* forces the specification of the actual properties, many of which are surface activity, gelling or membrane-forming, thus targeting whatever possible end-application in a clearer, direct, and faster manner. In this regard, most microbial glycolipids (e.g., sophorolipids, rhamnolipids, cellobioselipids) have a double amphiphilic, referred to as bolaform, molecular structure. In the literature, such molecules are historically defined as bolaamphiphiles, the solution self-assembly properties of which are also quite atypical when compared to classical amphiphiles.[76] Interestingly, the self-assembly of microbial glycolipids was discussed before within the context of bolaamphiphiles.[77]

It goes without saying that the proposition above has to face at least three decades of the use of the term "*biosurfactant*" in hundreds of research and review articles. It is then obviously unrealistic to achieve a change in the terminology in the immediate future. However, it is still important to raise the debate, broaden the perspectives and share the idea that this class of microbially-produced amphiphiles are not just surfactants, but they are a whole new class of molecules, some of which with unique all-in-one properties (e.g., glucolipid C18:1)[52] when compared to the wide range of existing natural and synthetic amphiphiles. It is then worth studying microbial bioamphiphiles not only in the market perspective of replacing synthetic chemicals with bio-based chemicals, but also of pushing forward the fundamental knowledge in the field of colloids and interface science.

**Conclusions**

Microbially-produced glycolipid and lipopeptide mixtures have historically been addressed as biosurfactants, for their ability to lower the surface tension of water, to form micelles and generate foam and oil-water emulsions. However, recent research in the field of physical chemistry applied to this class of compounds, especially to newer (genetically or chemically) engineered single molecules, shows a more complex behaviour, in particular by their neutral acidic form, thus questioning/assessing the use of the term *surfactant*. Just like phospholipids, bile salts and other complex organic acids (e.g., glycyrrhizic acid), microbial glycolipids and lipopeptides assemble into uni- and multilamellar vesicles and also into semicrystalline fibres, with sometimes unexpected gelling properties. This behaviour does not match the traditional definition of a surfactant, known to lower the surface tension of water but also to form micelles in a dynamic equilibrium beyond the CMC. Their behaviour in solution is then much richer and, by the systematic employment of the term *surfactant*, is underestimated. For this reason, and with the goal of improving the communication across communities and to broaden the application potential, this communication proposes to employ a more general expression describing these molecules as *microbial bioamphiphiles*. The expression *biological amphiphiles* was employed already twenty years ago by Prof. S. Lang,[4] therefore the goal of this commentary is to broaden the perspectives in using the word *amphiphile*, much more appropriately to molecules which have a surface activity, but also gelling and membrane-forming abilities.

**Acknowledgements**

My warmest acknowledgements to Dr. S. Roelants, Prof. I. Van Bogaert, Prof. W. Soetaert, Prof. C. Stevens at Gent University (Belgium), S. Ziebek at Fraunhofer Institute, Stuttgart (Germany), T. Tiso at RWTH Aachen (Germany), B. L.V. Prasad at NCL, Pune (India) for the fruitful discussions and providing me with microbial bioamphiphiles across the years. Acknowledgements also to Prof. Gloria Sobéron-Chavez (Universidad Autonoma de Mexico, Mexico) for commenting this work. A warm and thoughtful acknowledgement to Prof. I. M. Banat (Ulster University, UK) for reading, correcting and commenting this work. I am also very grateful to Prof. R. Hausmann (University of Hohenheim, Germany), Prof. Gloria Sobéron-Chavez and Prof. E. Deziel (University of Laval, Canada) for accepting this comment at the 2[nd] international Biosurfactant Conference held at the University of Hohenheim, Germany. I also thank the Reviewers of this manuscript, who strongly contributed to its present form.

**References highlight**

*This article is published within the context of the Special Issue on Biosurfactants at Current Opinion in Colloid and Surface Science*

Ref. 6 -**- Fungal Biosurfactants, from Nature to Biotechnological Product: Bioprospection, Production and Potential Applications –

*A recent review article in which biosurfactants are discussed from both a biological and functional perspective, with an emphasis on the techniques of characterization of the surfactant properties.*

---------------------

Ref. 18 - **- Biosurfactants: Opportunities for the development of a sustainable future -

*Authors critically review the field of biosurfactants in terms of their Strengths and opportunities but also their weaknesses and Threats.*[18]

--------------------

Ref. 22 - **- From Lab to Market: An Integrated Bioprocess Design Approach for New-to-Nature Biosurfactants Produced by Starmerella Bombicola

*A complete work spanning from genetic engineering giving rise to new sophorolipid derivatives to the evaluation of the application potential, passing through self-assembly*

---------------------

Ref. 23 -**- Starmerella Bombicola: Recent Advances on Sophorolipid Production and Prospects of Waste Stream Utilization -[23]

*Authors review the recent advances in the sophorolipid production in view of their commercialization. Literature is analysed from the point of view of recent achievements in metabolic engineering, bioprocessing and employment of renewable feedstock*

----------------------

Ref. 24 -**- Development of a Cradle-to-Grave Approach for Acetylated Acidic Sophorolipid Biosurfactants

*Probably the first multidisciplinary and multicollaborative work spanning from genetic engineering to life cycle analysis of a sophorolipid congener. This work also studies self-assembly and it correlates it to bulk properties like emulsification and foaming.*

--------------------

Ref. 31 -*- Glycyrrhizic Acid: Self-Assembly and Applications in Multiphase Food Systems -

*This article is published within the context of the Special Issue on Biosurfactants at Current Opinion in Colloid and Surface Science*

*Authors review the self-assembly and applications in O/W emulsions, gel emulsions, foams of glycyrrhizic acid, a biobased glycolipid saponin similar in structure to biosurfactants*[31]

---------------------

Ref. 38 -**- Solution Self-Assembly and Adsorption at the Air-Water Interface of the Monorhamnose and Dirhamnose Rhamnolipids and Their Mixtures -

*Authors study the self-assembly properties of rhamnolipids in bulk water and at interfaces using neutron scattering and reflectivity and they identify important structural parameters like area-per-molecule, micellar radius, aggregation number.* [38]

---------------------

Ref. 40 -*- Aggregation of the Naturally Occurring Lipopeptide, Surfactin, at Interfaces and in Solution: An Unusual Type of Surfactant?

*Using neutron reflectivity and scattering data recorded on deuterium-modified surfactin biosurfactant, authors propose a new model of surfactin aggregation.*[40]

--------------------

Ref. 52 -**- Chameleonic Amphiphile: The Unique Multiple Self-Assembly Properties of a Natural Glycolipid in Excess of Water

*Authors show that a so-called glycolipid biosurfactant in water is able to assemble into a broad range of morphologies (micelles, fibers, membranes, giant micelles, multilamellar wall vesicles, onions…) within relatively restraint physicochemical conditions.*[52]

----------------------

Ref. 55 -**- Metallogels from Glycolipid Biosurfactant.

*The first study demonstrating metallogelation of biosurfactants and combining rheology, SAXS and cryo-TEM*

--------------------

Ref. 66 -**- Solution Self-Assembly of the Sophorolipid Biosurfactant and Its Mixture with Anionic Surfactant Sodium Dodecyl Benzene Sulfonate

*Authors use neutron scattering to study the self-assembly of lactonic and acidic sophorolipids and their mixture. They suppose vesicular structures of lactonic sophorolipids.*

--------------------

*This article is published within the context of the Special Issue on Biosurfactants at Current Opinion in Colloid and Surface Science*

Ref. 67 -**- Natural Synergism of Acid and Lactone Type Mixed Sophorolipids in Interfacial Activities and Cytotoxicities

*This work suggests that the natural mixture of lactonic and acidic sophorolipids have improved interfacial properties than the single acidic and lactonic forms. This could help understanding the reason why wild type microorganisms produce mixtures of metabolites instead of pure compounds.*

---------------------

Ref. 72 -**- Unlocking the Potential of a Glycolipid Platform through Chemical Modification

*Authors hypothesize that combination of genetic and chemical engineering is the strategy to develop new biobased performing bioamphophiles.*[72]

---------------------

Ref. 76 -**- Unlocking the Potential of a Glycolipid Platform through Chemical Modification

*An historical document illustrating the astonishing properties of amphiphiles with bolaform molecular structure.*[76]

---------------------

*This article is published within the context of the Special Issue on Biosurfactants at Current Opinion in Colloid and Surface Science*